\documentclass[11pt]{article}

\usepackage{amsmath,amssymb}
\usepackage{graphicx}
\usepackage{epstopdf}
\usepackage{color}

\newcommand{\be}{\begin{equation}}
\newcommand{\ee}{\end{equation}}
\newcommand{\bea}{\begin{eqnarray}}
\newcommand{\eea}{\end{eqnarray}}
\newcommand{\ba}{\begin{array}}
\newcommand{\ea}{\end{array}}

\numberwithin{equation}{section}

\newcommand{\vev}[1]{\left< #1 \right> }
\newcommand{\nn}{\nonumber}

\newcommand{\vevs}[1]{\langle #1 \rangle }
\newcommand{\sket}[1]{| #1 \rangle}
\newcommand{\sbra}[1]{\langle  #1 |}
\newcommand{\sbracket}[2]{\langle #1 |  #2 \rangle}

\begin{document}

\allowdisplaybreaks

\title{Entropic uncertainty relation \\ based on generalized uncertainty principle \bigskip}

\author{Li-Yi Hsu${}^1$, Shoichi Kawamoto${}^{1,2}$, and Wen-Yu Wen${}^{1,2,3}$
	\\
${}^1$\small\it  Department of Physics, 
 Chung Yuan Christian University, Taoyuan City, Taiwan\\
${}^2$\small\it  Center for High Energy Physics,
 Chung Yuan Christian University, Taoyuan City, Taiwan\\
${}^3$\small\it  Leung Center for Cosmology and Particle Astrophysics,
National Taiwan University, Taipei 106, Taiwan\\
\footnotesize\tt 
lyshu, kawamoto, wenw@cycu.edu.tw
	}

\date{\today}

\maketitle

\bigskip

\begin{abstract}
\noindent\normalsize
We explore the modification of the entropic formulation of uncertainty
principle in quantum mechanics which measures the incompatibility of
measurements in terms of Shannon entropy. The deformation in question
is the type so called generalized uncertainty principle that is
motivated by thought experiments in quantum gravity and string theory
and is characterized by a parameter of Planck scale. The corrections
are evaluated for small deformation parameters by use of the Gaussian
wave function and numerical calculation. As the generalized
uncertainty principle has proven to be useful in the study of the
quantum nature of black holes, this study would be a step toward
introducing an information theory viewpoint to black hole physics.

\end{abstract}

\vfill

\section{\label{sec:intro} INTRODUCTION}

Heisenberg's Uncertainty Principle (HUP) is one of the most
fundamental relations in quantum physics \cite{aHeisenberg:1927zz}.
It is usually stated in Robertson's form \cite{Robertson:1929zz},
\begin{align}
 \Delta_\psi A \Delta_\psi B \geq \frac{1}{2} | \langle [A,B] \rangle_\psi | \,,
\end{align}
where $\vevs{\cdots}_\psi$ is the expectation value with respect to the state $\psi$ and
$\Delta_\psi X = \sqrt{\vevs{X^2}_\psi-\vevs{X}_\psi^2}$ is the standard deviation of an observable $X$ with respect to $\psi$.
This relation characterizes the complementarity of quantum mechanics; for an incompatible pair
of measurements it is impossible to carry out precise measurement simultaneously.
It should be noted that the lower bound is state dependent and occasionally the bound becomes meaningless; for example, if the state is an eigenstate of either of the observables.
Since uncertainty relation is expected to characterize the
incompatibility of measurements, 
it is thus also important to give a state-independent bound \cite{Deutsch}.
Along this line, several extensions of the relation have been proposed.
Among them, entropic uncertainty relation (EUR) serves a
state-independent notion of uncertainty that is characterized by
Shannon entropy of the probability distribution as a measure of the
imperfection of knowledge.
For a finite dimensional case, Maassen and Uffink \cite{MU1}
gave a class of EUR based on a conjecture by Kraus \cite{Kraus},
where the most stringent bound
is associated with a 
mutually unbiased pair of observables.
For continuum observables, such as the position and momentum which we are interested in,
Bia\l ynicki-Birula and Mycielski \cite{BB-Myceielski} 
proposed an EUR based on a mathematical inequality for Fourier transform \cite{Babenko,Beckner}.
Later this uncertainty relation has been refined by taking the resolution of measurement into account
\cite{BB1, BB-Eudnicki}.

On the other hand, it has been suspected that Heisenberg's uncertainty relation needs to be
modified if quantum gravity effect becomes important.
In quantum gravity, the space-time itself is subjected to quantum fluctuation and the uncertainty of position measurement is enhanced by further quantum fluctuation.
For instance, if we attempt to detect the position of a particle with a light source of ultra high frequency in order to achieve high precision,
the energy of the photon will perturb the background geometry and we will not achieve desired
accuracy \cite{Garay:1994en}.
This argument has been made more precise, through, for example, the analysis of high energy scattering in string theory \cite{Yoneya:2000bt} or the analysis of black holes\cite{Maggiore:1993rv,Adler:2001vs, Fabio1}
, and the result is encapsulated in the so-called generalized uncertainty principle (GUP).
GUP can be regarded as a deformation of HUP with a scale set by the Planck length
and is considered to be a phenomenological realization of this extra uncertainty .

It is then natural to ask how EUR should be modified if we consider a quantum gravity effect.
In this paper, we explore the correction to EUR for position and momentum observables by taking
GUP into account.
GUP can be realized as a deformed commutation relation of the position and the momentum
which has a Hilbert space representation \cite{Kempf:1994su}.
With this representation, we may spot the part of wave functions due to GUP in the EUR.
To be specified, we start with Gaussian wave functions which saturates EUR bound for the undeformed case.
It is thus the deviation from the undeformed case can be considered to well approximate
the possible correction terms that come from the  GUP effect.

This paper is organized as follows: In section \ref{sec:EUR}, some overview of EUR is presented.
Section \ref{sec:gener-uncert-princ} introduces GUP and deformed wave functions.
In section \ref{sec:gener-uncert-corr}, we consider EUR by use of deformed wave functions
and explore the corrections to EUR due to GUP.
Finally, conclusion and discussion is given in section \ref{sec:conclusion}.

\section{\label{sec:EUR}Entropic uncertainty relation}
\label{sec:entr-uncert-relat}

\begin{figure}[h]
  \centering
  \includegraphics[scale=0.8]{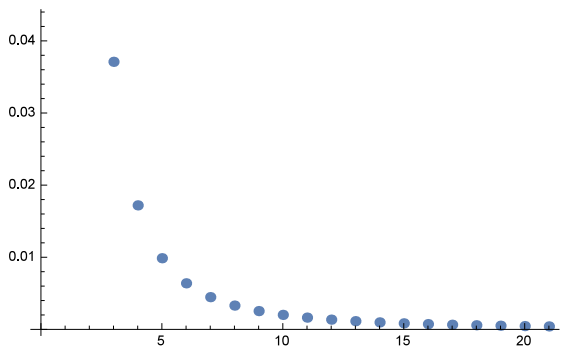}
  \caption{The difference between the left hand side and the right hand side of \eqref{eq:EUR1} for Gaussian wave functions \eqref{eq:Gaussian_wf0} (with $\hbar=1$). The horizontal label corresponds to $N=10 \times n +1$.  \label{fig:EUR_Gaussian1}}
\end{figure}

Entropic formulation of uncertainty relation is written as
\begin{align}
  \label{eq:1}
  H(A| \psi) + H(B| \psi)  \geq c(A,B)
\end{align}
where $H(A|\psi)$ is Shannon entropy of the probability distribution of the observable
$A$ with respect to a quantum state $|\psi\rangle$, and $c(A,B)$ is a state-independent bound.
Since Shannon entropy represents ignorance on the distribution, this relation measures an incompatibility of the measurement of $A$ and $B$.
According to various choices of observables and systems, different types of EUR have been proposed.
For a finite dimensional case, we consider observables that have the following spectral decompositions,
$A=\sum_i a_i P_{a_i}$ and $B= \sum_j b_j P_{b_j}$ where $P_{a_i}$ and $P_{b_j}$ are projection
operators.
The Shannon entropy is defined by $H(A|\psi)= - \sum_i p_i^{(a)} \log p_i^{(a)}$
with $p_i^{(a)}=\langle \psi | P_{a_i} | \psi \rangle$
and  $H(B|\psi)$ is also defined in a similar manner.\footnote{%
Since the Shannon entropy is a convex function, the same inequality holds for a general quantum state
$\hat{\rho}=\sum_i c_i |\psi_i \rangle \langle \psi_i|$ with $\sum_i c_i=1$.}
Maassen and Uffink showed that EUR \eqref{eq:1} holds with $c(A,B)=-2 \log ( \max_{i,j} | \langle a_i | b_j \rangle |) $ \cite{MU1}.
It is then the largest lower bound is realized when the inner products of each eigenbasis
are $1/\sqrt{d}$ with $d$ being the dimension of the Hilbert space.
For observables with continuum spectra, $\hat{x}$ and $\hat{p}$, 
Bia\l ynicki-Birula and Mycielski gave the bound,
\begin{align}
 \tilde{H}(x| \psi) + \tilde{H}(p| \psi) \geq \log (e\pi)  \,,
\end{align}
where 
$\tilde{H}(x| \psi ) = - \int dx \rho(x) \log \rho(x)=-\vev{\log \rho(x)}_\rho$ and 
$\rho(x)=|\psi(x)|^2$ is the probability density of the position.  
Similar definition is for $\tilde{H}(p|\psi)$ \cite{BB-Myceielski}.
Later, this bound is refined as \cite{BB1}
\begin{align}
\label{eq:EUR1}
  H(x|\psi) + H(p|\psi) \geq \log \bigg( \frac{\pi e \hbar}{\delta x \delta p} \bigg) \,,
\end{align}
where $\delta x$ and $\delta p$ are the resolution of the position and the momentum measurement
respectively.
Namely, the position (momentum) space is divided into many bins with the width $\delta x$
($\delta p$),
and $p_i^x$ ($p^p_j$) is the probability of finding a particle at the $i$-th ($j$-th) bin,
\begin{align}
  p_i^x = \int_{i \delta x}^{(i+1) \delta x} dx \rho(x) \,.
\end{align}
Similarly, $p_j^p = \int_{j \delta p}^{(j+1) \delta p} dp \rho(p)$ for momentum space.  
Then the Shannon entropies are calculated via $H(x|\psi) = -\sum_i p_i^x \log p_i^x$ and
$H(p|\psi)=-\sum_j p_j^p \log p_j^p$.
This refined Shannon entropy satisfies the inequality
$H(x|\psi) \geq \tilde{H}(x|\pi) - \ln \delta x$ (the same for $H(p|\psi)$),
and \eqref{eq:EUR1} follows.
This form nicely reflects the fact that the actual measurements for position and momentum are restricted by the resolution of measurement apparatus.
The bound is not significant if the resolution is {\sl classical}, namely $\delta x \delta p \gg \hbar$, while it tends to give tight bounds when we consider fine measurements
$\delta x \delta p \simeq \hbar$.
In particular, the bound is saturated by Gaussian wave functions
\begin{align}
  \psi_0(x) = \bigg( \frac{\hbar^2}{\pi} \bigg)^{1/4} e^{-\frac{\hbar^2}{2}x^2}\,,
\quad
  \tilde{\psi}_0(p) =
\bigg( \frac{1}{\pi \hbar^2}\bigg)^{1/4} e^{-\frac{p^2}{2\hbar^2}} \,,
\label{eq:Gaussian_wf0}
\end{align}
with fine bin sizes, $\delta x \delta p \ll \hbar$.
The difference between the left hand side and the right hand side of \eqref{eq:EUR1}
with respect to the bin size
is plotted in Fig.~\ref{fig:EUR_Gaussian1}.
Here, the region $x,p \in [-50,50]$ is divided in to $2N$ bins, and then the bin size is
$\delta x, \delta p = 50/N$.

\section{\label{sec:gener-uncert-princ}Generalized uncertainty principle}

As described in Sec \ref{sec:intro}, GUP is a deformation of the standard Heisenberg
uncertainty principle.
In general, GUP is characterized by the relation,
\begin{align}
  \Delta x \Delta p
 \geq \frac{\hbar}{2} \big(  1 + \alpha (\Delta x)^2
+ \beta (\Delta p)^2 + \gamma \big) \,,
\end{align}
where $\alpha$, $\beta$, and $\gamma$ are positive deformation parameters.
They are independent of $\Delta x$ and $\Delta p$, but may depend on the expectation values
$\langle \hat{x} \rangle$ and $\langle \hat{p} \rangle$.
GUP is motivated by gedanken experiments in quantum gravity,
such as loop quantum gravity, string theory, and black hole physics,
and in those cases GUP is represented by a restricted relation,
\begin{align}
  \Delta x \Delta p
 \geq \frac{\hbar}{2} \big(  1 
+ \beta (\Delta p)^2 + \gamma \big) \,,
\label{eq:GUP1}
\end{align}
and the values of the parameters depend on the characteristic scale of
underlying theory of gravity.
This form can be understood as the result of a standard Robertson uncertainty relation,
\begin{align}
  \Delta A \Delta B \geq \frac{1}{2} \big| \langle [A,B] \rangle \big|
\,,
\end{align}
in which the canonical commutation relation is deformed,
\begin{align}
  [\hat{x}, \hat{p} ] = i\hbar \big( 1+ \beta \hat{p}^2 \big) \,,
\end{align}
and $\gamma= \beta \langle \hat{p} \rangle^2$ is chosen.

The minimum uncertainty in $x$ is $\Delta x_\text{min}=\hbar \sqrt{\beta}$
which is realized for $\vev{p}=0$ states (thus $\gamma=0$).
This corresponds to the case with $\Delta p= \beta^{-1/2}$ 
and $\Delta x_\text{min} \Delta p = \hbar$.
Hence there does not exist the eigenstate of the position operator $\hat{x}$;
not even an approximated normalizable state.
We thus work in the momentum space, $\hat{p} |p\rangle =  p |p\rangle$.
The Hilbert space representation is constructed in \cite{Kempf:1994su}.
On the momentum space wave function $\psi(p)= \langle p | \psi \rangle$, 
the operators are realized as $\hat{p} \psi(p)=p \psi(p)$
and $\hat{x} \psi(p)= i \hbar (1+\beta p^2) \partial_p \psi(p)$.
In order for $\hat{x}$ and $\hat{p}$ operators to be symmetric, a momentum factor is involved in the definition of scalar product and the completeness condition, that is
\begin{align}
\label{eq:inner_prod_GUP}
  \sbracket{\psi}{\phi} = \int_{-\infty}^\infty \frac{dp}{1+\beta p^2} \psi^*(p) \phi(p)
\,,
\end{align}
and 
\begin{align}
 1=\int \frac{dp}{1+\beta p^2} \sket{p}\sbra{p}
\,.
\end{align}

\subsection{Gaussian wave function in momentum space}
\label{sec:gauss-wave-funct}

We start with the Gaussian wave function in momentum space $\tilde{\psi}_0(p)$ in
\eqref{eq:Gaussian_wf0} for undeformed ($\beta=0$) case.
The expectation value of $p$ is $\vev{p}=0$ and the standard deviation is
$\Delta p = \sqrt{\vev{(p-\vev{p})^2}} = {\hbar}/{\sqrt{2}}$.
Now we consider the GUP case and define a ``position-space'' wave function.
Before we proceed, it should be noted that the momentum distribution function
is defined as $\rho(p)=|\psi_0(p)|^2/(1+\beta p^2)$ due to the normalization condition
$\sbracket{\psi_0}{\psi_0}=1$ in \eqref{eq:inner_prod_GUP}.
Namely, with GUP, the momentum space wave function is Gaussian but the
distribution is not.

For a normalized state $\sbracket{\psi}{\psi}=1$, the corresponding momentum space
Gaussian wave function with GUP reads
\begin{align}
  \tilde{\psi}_\beta(p) =& \sbracket{p}{\psi}
= F(\beta) e^{-\frac{\sigma^2 p^2}{2\hbar^2}} \,,
\end{align}
where $\sigma$ is a parameter for the width (more precisely, the half width is given by
$2\sqrt{2\ln 2}\hbar /\sigma$), and
$F(\beta)$ is the normalization factor,
\begin{align}
  F(\beta)=&
e^{-\frac{\sigma^2}{2 \hbar^2 \beta}} \pi^{-1/2} \beta^{1/4} 
\bigg[ \mathrm{erfc} \bigg(\frac{\sigma}{\hbar \sqrt{\beta}} \bigg)  \bigg]^{-1/2} \,,
\end{align}
and $\mathrm{erfc}(z)= \frac{2}{\sqrt{\pi}}\int_{z}^\infty dt \, e^{-t^2}$ is the complementary error function.
Note that $F(\beta)$ is monotonically increasing in $\beta$ and
$F(0)=(\sigma^2/\pi \hbar^2)^{1/4}$.
We then define the probability distribution of momentum as
\begin{align}
  \rho_\beta (p) =& \frac{| \tilde{\psi}_\beta(p) |^2}{1+\beta p^2}
 \,,\qquad
\int_{-\infty}^\infty dp \, |\rho_\beta(p)|^2=1 \,.
\label{eq:rho_beta_p}
\end{align}
With this probability distribution, the average $\vev{p}=0$ remains but the deviation $\Delta p$ is altered.
This will affect the position uncertainty as we shall see later.

\subsection{Pseudo-position space wave function}
\label{sec:pseudo-posit-space}

In \cite{Kempf:1994su}, a pseudo-position eigenstate
$\sket{\psi_\xi^\text{ML}}$ is constructed
and its
momentum space representation is
\begin{align}
    \psi^\text{ML}_\xi(p) =& 
 \sqrt{\frac{2\sqrt{\beta}}{\pi}}
(1+\beta p^2)^{-1/2} \exp \bigg[
- i \frac{\xi \arctan (\sqrt{\beta} p)}{\hbar \sqrt{\beta}}
\bigg]  \,.
\end{align}
This state has minimum uncertainty,
$\vev{\hat{x}}=\xi$ and $\Delta x = \Delta  x_\text{min}=\hbar
\sqrt{\beta}$.
By using this, a position space wave function is obtained from
a momentum space one as
\begin{align}
  \psi(\xi)= 
 \sqrt{\frac{2\sqrt{\beta}}{\pi}}
\int_{-\infty}^\infty \frac{\tilde{\psi}_\beta(p) dp}{(1+\beta p^2)^{3/2}}
 \exp \bigg[
i \frac{\xi \arctan (\sqrt{\beta} p)}{\hbar \sqrt{\beta}}
\bigg]   \,.
\label{eq:position_wf_KMM}
\end{align}
For a suitably normalized momentum space wave function $\int
\frac{dp}{1+\beta p^2} |\tilde{\psi}(p)|^2=1$, this $\psi(\xi)$
is not normalized, instead
\begin{align}
  \int_{-\infty}^\infty d\xi \, |\psi(\xi)|^2=
4\sqrt{\beta} \hbar
 \int_{-\infty}^\infty \frac{|\psi(p)|^2 dp}{(1+\beta p^2)^2} \,.
\end{align}
Namely, $\xi$ integration (or the completeness of $\sket{\psi_\xi^\text{ML}}$)
needs to involve a $p$ dependent factor
$(4\sqrt{\beta} \hbar)^{-1} (1+\beta p^2)$.
In order to define the probability distribution,
we include this factor
and use, instead of \eqref{eq:position_wf_KMM},
 the following pseudo-position space wave function\footnote{%
As we discuss in Section \ref{sec:using-auxil-moment}, this wave function is
related to a wave function in an auxiliary momentum space.}
\begin{align}
    \psi_\beta(\xi) =& 
\frac{1}{\sqrt{2\pi \hbar}}
\int_{-\infty}^\infty
\frac{\tilde{\psi}_\beta(p) dp}{1+\beta p^2} \exp \bigg[
i \frac{\xi \arctan (\sqrt{\beta} p)}{\hbar \sqrt{\beta}}
\bigg]
  \,.
\label{eq:xi_wf}
\end{align}
We define the probability distribution in
$\xi$ space as
\begin{align}
  \rho_\beta(\xi)=& |\psi_\beta(\xi)|^2 \,,
\qquad
\int_{-\infty}^\infty d\xi \, \rho_\beta(\xi) =1 \,.
\label{eq:rho_beta_xi}
\end{align}

\begin{figure}[h]
  \centering
  \includegraphics[scale=0.6]{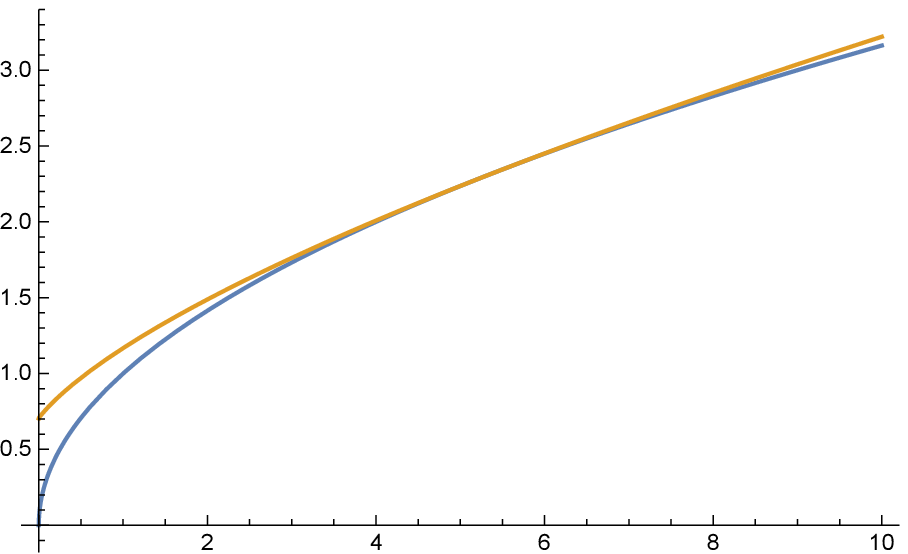}
\hspace{1em}
  \includegraphics[scale=0.6]{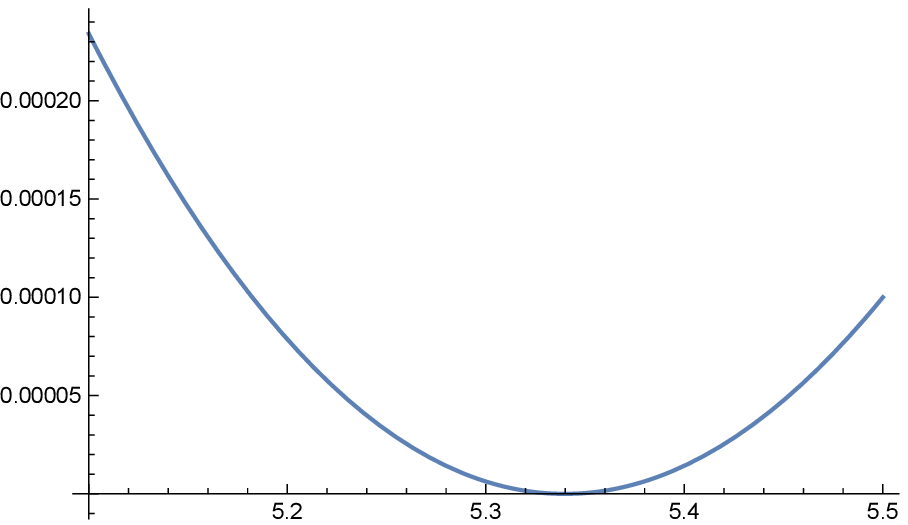}
  \caption{
The lower bound of the uncertainty in $x$ ($\hbar=\sigma=1$).
The horizontal axis is $\beta$.
[Left] $\Delta x_\text{min}=\hbar \sqrt{\beta}$ (blue)
and $\Delta x_\text{min, Gaussian}$ (red).
[Right] The plot for
$\Delta x_\text{min, Gaussian} - \Delta x_\text{min} $
with respect to $\beta$.
 \label{fig:min_uncertainty}}
\end{figure}

We next examine the lower bound of the uncertainty in the pseudo-position space
$\xi$ for the Gaussian wave function.
In the momentum space, the deviation for $\tilde{\psi}_\beta$ is
\begin{align}
\Delta p = \frac{1}{\sqrt{\beta}}
\sqrt{ -1 + \frac{\hbar \sqrt{\pi}}{\sigma} F(\beta)^2 } \,.
\label{eq:position_unc_1}
\end{align}
From GUP, the uncertainty for Gaussian wave function reads
$ \Delta x_\text{min, Gaussian} =
\frac{\hbar}{2 \Delta p} \big( 1 + \beta (\Delta p)^2  \big)$.
$\beta$ dependence of the minimum value, while the inequality is saturated,
is shown in Fig.~\ref{fig:min_uncertainty}.

\section{Generalized Uncertainty correction to entropic uncertainty relation}
\label{sec:gener-uncert-corr}

In this section, we explore corrections to the entropic uncertainty
relation due to GUP for the case of the Gaussian wave function.

The entropic uncertainty relation is based on the following probability distributions
and the Shannon entropies,
\begin{align}
  \label{eq:21}
&  H(p|\psi) =- \sum_i p_i \ln p_i \,,
\qquad
p_i = \int_{i}^{i+\delta p} dp \, {\rho}_\beta(p) \,,
\\
&  H(\xi|\psi) =- \sum_i q_i \ln q_i \,,
\qquad
q_i = \int_{i}^{i+\delta \xi} d\xi \, \rho_\beta(\xi) \,,
\end{align}
where $\rho_\beta(p)$ and $\rho_\beta(\xi)$ are given by
\eqref{eq:rho_beta_p} and \eqref{eq:rho_beta_xi} respectively.
For sufficiently small $\beta$, the sum of these entropies will be
bounded by the standard bound with a correction term depending on $\beta$,
\begin{align}
  \label{eq:3}
  H(\xi | \psi) + H(p | \psi)\geq \log \bigg( \frac{\pi e \hbar}{\delta x \delta p} \bigg)
+K(\beta; \delta \xi,\delta p) \,.
\end{align}
The right hand side is expected to be a state independent bound.
Since in the undeformed case ($\beta=0$) the Gaussian wave function is to saturate
the bound for sufficiently small bin sizes,
we expect that the
difference of the left hand side from the undeformed result approximates the correction
function $K(\beta;\delta \xi, \delta p)$,
\begin{align}
  \label{eq:6}
H_\text{tot}  -H_\text{tot}|_{\beta=0}
\simeq K(\beta; \delta \xi,\delta p) \,,
\end{align}
where $H_\text{tot} = H(\xi | \psi) + H(p | \psi)$.
As we have seen, the difference is almost saturated around $\delta \xi= \delta p=0.5$
for the undeformed case,
we may fit the correction function for small values of $\beta$.

\begin{figure}[tbh]
  \centering
  \includegraphics[scale=0.6]{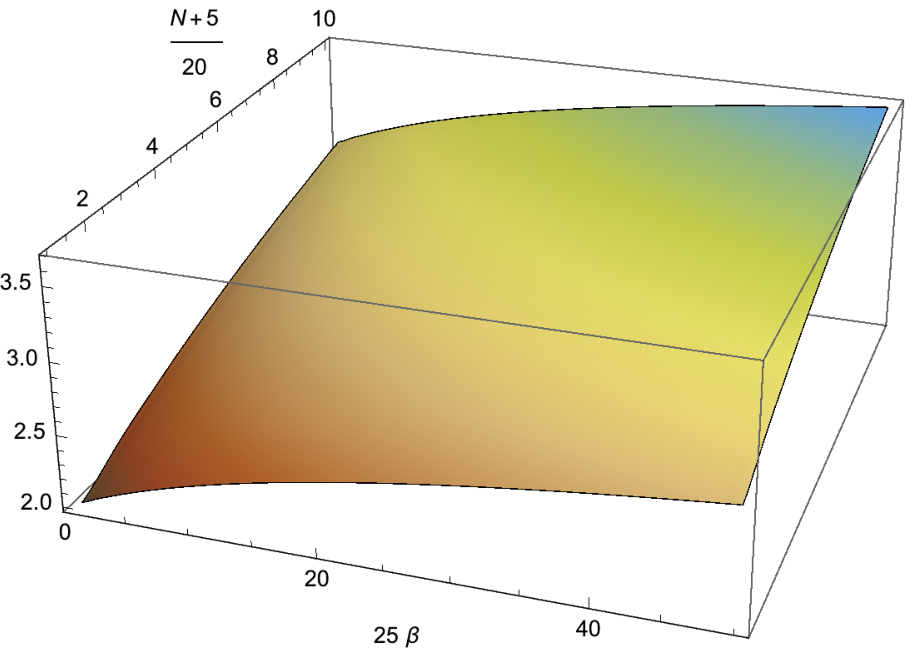}
\hspace{1em}  
  \includegraphics[scale=0.6]{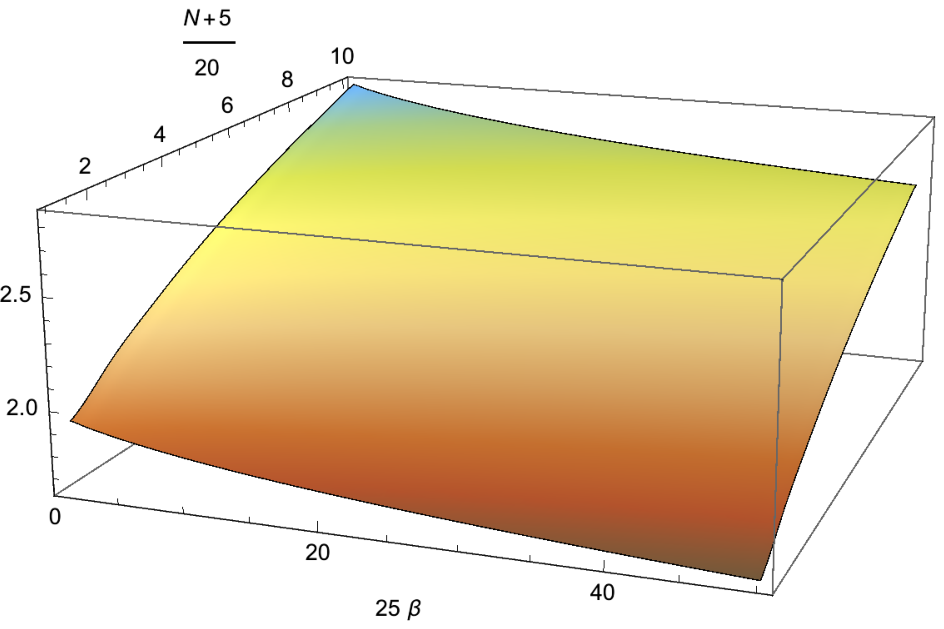}
  \caption{The $\beta$-deformed Shannon entropy.
Left is $H(\xi|\psi)$ and the right is $H(p|\psi)$.
$N$ is from $10$ to $100$ by step $10$.
$\beta$ is from $1/25=0.04$ to $50/25=2$ by step $0.04$.} 
\label{fig:shanonn_3D}
\end{figure}

These $\beta$-deformed Shannon entropy of the position and the momentum are plotted
for various $\beta$ and bin sizes
in Fig.~\ref{fig:shanonn_3D}.
Here, the interval from $-50$ to $50$ is divided into $2N$ bins.
Thus, the bin size $\delta \xi$ and $\delta p$ is $50/N$.

\begin{figure}[tbh]
  \centering
  \includegraphics[scale=0.5]{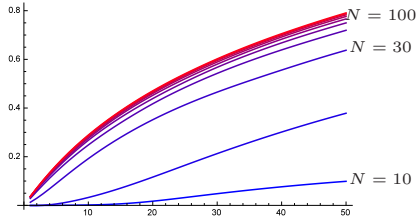}
\put(10,15){\makebox(0,0){\tiny $N=10$}}
\put(10,65){\makebox(0,0){\tiny $N=30$}}
\put(10,77){\makebox(0,0){\tiny $N=100$}}
  \caption{$H(\xi|\psi)-H(\xi|\psi)|_{\beta=0}$ against $\beta$
for $N=10$ (dark blue) to $N=100$ (red).
The horizontal axis is $\beta/25$.
\label{fig:difference_shanonnXi}} 
\end{figure}

\begin{figure}[tbh]
  \centering
  \includegraphics[scale=0.5]{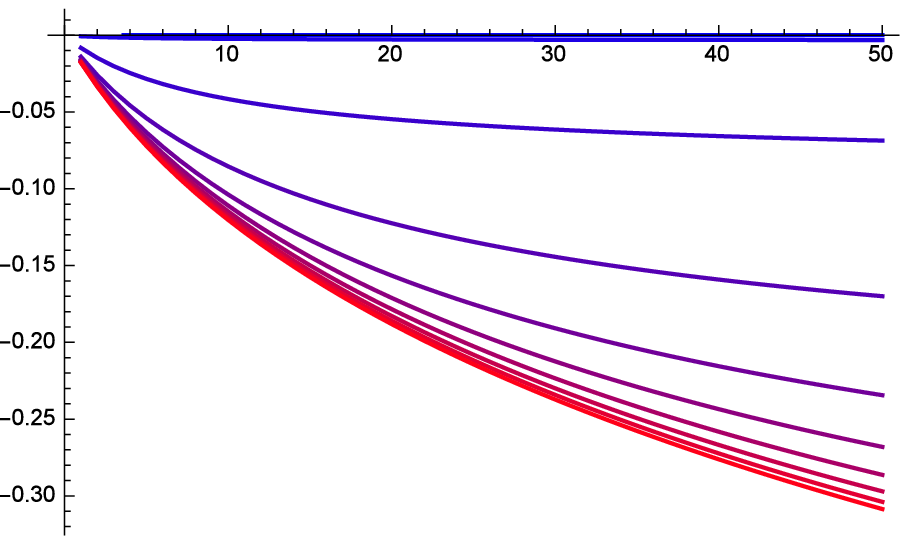}
\put(10,72){\makebox(0,0){\tiny $N=10$}}
\put(10,35){\makebox(0,0){\tiny $N=30$}}
\put(10,3){\makebox(0,0){\tiny $N=100$}}
  \caption{$H(p|\psi)-H(p|\psi)|_{\beta=0}$ against $\beta$
for $N=10$ (dark blue) to $N=100$ (red).
The horizontal axis is $\beta/25$.
\label{fig:difference_shanonnP}} 
\end{figure}

\begin{figure}[tbh]
  \centering
  \includegraphics[scale=0.5]{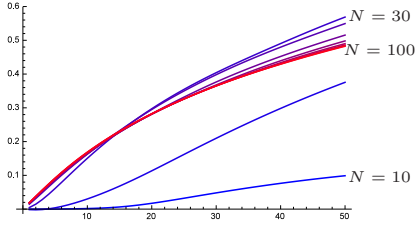}
\put(10,17){\makebox(0,0){\tiny $N=10$}}
\put(10,78){\makebox(0,0){\tiny $N=30$}}
\put(10,65){\makebox(0,0){\tiny $N=100$}}
  \caption{$H_\text{tot}-H_\text{tot}|_{\beta=0}$ against $\beta$
for $N=10$ (dark blue) to $N=100$ (red).
The horizontal axis is $\beta/25$.
\label{fig:difference_shanonnLHS}} 
\end{figure}

To see the correction due to $\beta$, we plot the difference 
of Shannon entropies between $\beta$-deformed and undeformed cases.
$H(\xi|\psi)-H(\xi|\psi)|_{\beta=0}$,
$H(p|\psi)-H(p|\psi)|_{\beta=0}$,
and $H_\text{tot}-H_\text{tot}|_{\beta=0}$
are shown in Fig.~\ref{fig:difference_shanonnXi}, 
Fig.~\ref{fig:difference_shanonnP}
and Fig.~\ref{fig:difference_shanonnLHS} respectively.
The horizontal axis represents $\beta/25$; namely plot region is
for $0 \leq \beta \leq 2.0$.
The bin size is chosen as $\delta x = \delta p = \frac{50}{N}$
and $N=10, 20, \cdots, 100$ are shown.
For small bin size, $N=80$, $90$,and $100$, the $\beta$ dependence tends to
be convergent (we have checked up to $N=300$ that the $N=100$ line is
sufficiently stable for $\beta \leq 2.0$).

The $\beta$ dependence appears in an opposite way between
the difference of Shannon entropy for the position operator
and that for the momentum operator.
The pseudo-position wave function $\psi_\beta$ increases
uncertainty in position as $\beta$ gets bigger, since its minimum position uncertainty
$\Delta x_\text{min, Gaussian}$ is an increasing function of $\beta$ as shown in 
Fig.~\ref{fig:min_uncertainty}.
On the other hand, in the momentum distribution function $\rho_\beta(p)=|\psi_0(p)|^2/(1+\beta p^2)$,
the $\beta$ dependence is only in the factor $1/(1+\beta p^2)$.
As $\beta$ increases, this factor tends to squeeze up $\rho(p)$ near $p=0$, and the uncertainty
gets smaller.
Therefore, it is natural that the Shannon entropy for the position operator is an 
increasing function of $\beta$ while that for the momentum operator is a decreasing one.
The sum, $H_\text{tot}=H(\xi | \psi) + H(p | \psi)$, is a monotonically increasing function of $\beta$.
This sum is to measure the incompatibility of the measurements of $x$ and $p$, and should be
related to $\Delta x \Delta p$.
For the $\beta$-deformed Gaussian wave functions,
$\Delta x \Delta p = \frac{\hbar^2 \sqrt{\pi}}{2\sigma}F(\beta)^2$.
$F(\beta)$ is a monotonically increasing function of $\beta$ and so is $\Delta x \Delta p$.
We thus again observe that the $\beta$ dependence is consistent with what we anticipate from the variance
type GUP observation.

\subsection{Evaluation of the correction term}
\label{sec:eval-corr-term}

Having obtained
the results for some values of $\beta$, we will fit the points
($N=100$ case) by a function of $\beta$.
Since the plot suggests a concave function, we may use a power function of $\beta$,
$\beta^b$, and the result is 
\begin{align}
  \label{eq:8}
K(\beta; 0.5,0.5) \sim
  \begin{cases}
    \beta^{0.772054} & (\beta \leq 1.0) \\
    \beta^{0.654778} & (\beta \leq 2.0) 
  \end{cases}
\,.
\end{align}
It turns out that, even for such small regions for $\beta$,
the fitting by a polynomial of $\beta$ (to tenth
order) does not give a stable result.

\begin{figure}[tbh]
  \centering
  \includegraphics[scale=0.6]{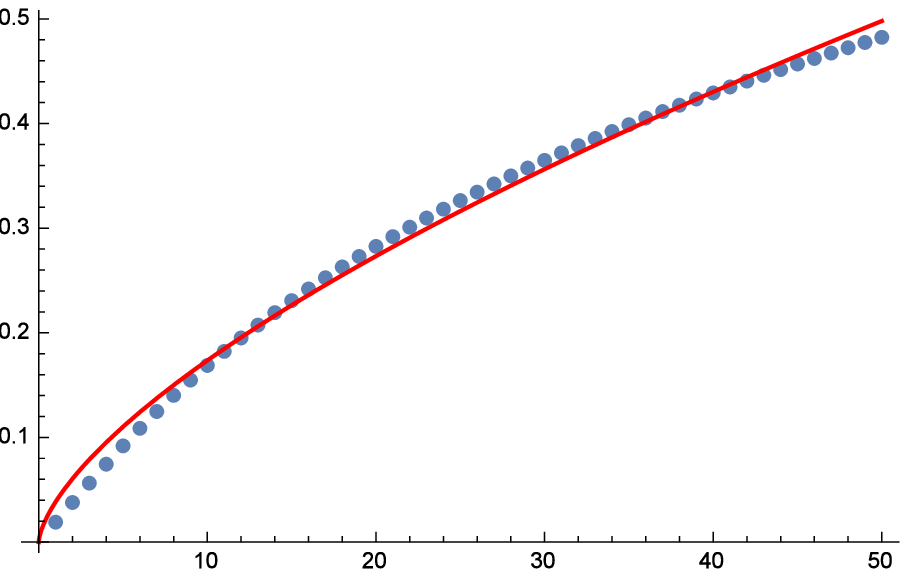}
\hspace{1em}
  \includegraphics[scale=0.6]{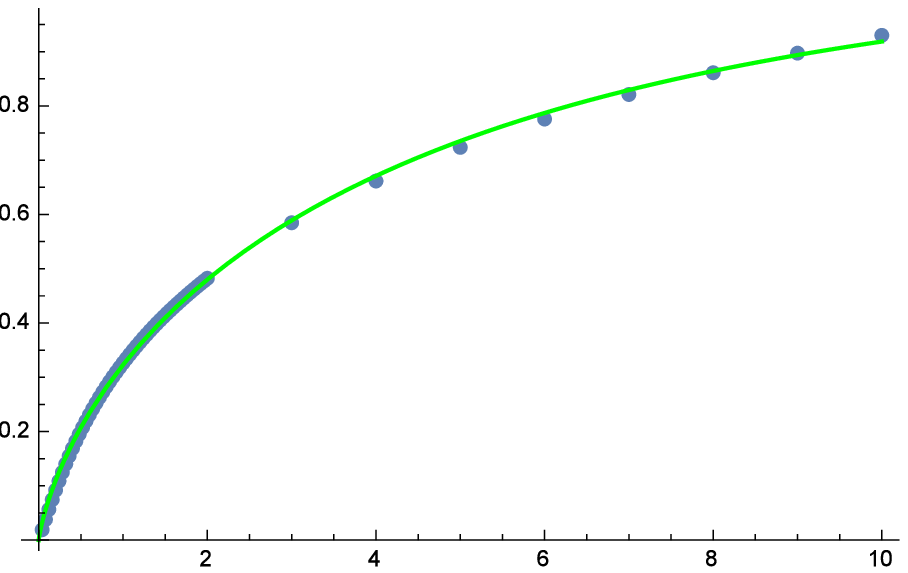}
  \caption{Fitting for the correction term $K(\beta; 0.5,0.5)$:
(Left) by $\beta^{0.654778}$ for $\beta \leq 2.0$.
(Right) $  -0.204103 \beta \ln \beta
+0.32293 \beta^{1.24049}$ against the data points $\beta \leq 10$.\label{fig:fitting1}}  
\end{figure}

We have also tried to fit a result with even larger $\beta$;
$0 \leq \beta \leq 10$ with $N=300$ and
the bin size $0.5$ (namely the integration region is taken $\xi, p \in
[-150, 150]$).
In this case, the fitting of a type $ a \beta \ln \beta  + b \beta^c$
turns out to be fairly good,
\begin{align}
  \label{eq:9}
K(\beta; 0.5,0.5) \simeq  -0.204103 \beta \ln \beta
+0.32293 \beta^{1.24049} \,.
\end{align}
These fittings are shown in Fig.~\ref{fig:fitting1}.
However, for large values of $\beta$ (say $\beta \geq 3.0$) it would
be necessary to use finer bin sizes to saturate the bound
since the data points may not be sufficiently close to the saturation of the bound.

\subsubsection{Using auxiliary momentum variable}
\label{sec:using-auxil-moment}

In the paper \cite{Pedram:2016gps}, the entropic uncertainty relation
has been investigated by use of an auxiliary momentum variable
\begin{align}
  p=\frac{1}{\sqrt{\beta}} \tan \big(\sqrt{\beta} q \big) \,,
\end{align}
where $x$ and $q$ satisfy the standard Heisenberg algebra, $[x,q]=i\hbar$.
Note that $\tilde{\psi}(q)$ is defined on a finite interval
$-q_0 \leq q \leq q_0$ with $q_0=\pi/2\sqrt{\beta}$.
Thus, the wave function $\psi(x)$ and $\tilde{\psi}(q)$ form the standard Fourier pair,
\begin{align}
  \psi(x)=& \frac{1}{\sqrt{2\pi \hbar}}\int_{-q_0}^{q_0} dq \, \psi(q) e^{\frac{ixq}{\hbar}} \,.
\end{align}
In terms of the physical momentum $p$, this relation reads
\begin{align}
  \psi(x)=& \frac{1}{\sqrt{2\pi \hbar}} \int_{-\infty}^\infty dp \,
\frac{\psi(p)}{1+\beta p^2} 
\exp \bigg[
i \frac{x \arctan (\sqrt{\beta} p)}{\hbar \sqrt{\beta}}
\bigg] \,,
\end{align}
where $\psi(p)=\psi(q)|_{q=\beta^{-1/2} \arctan (\sqrt{\beta}p)}$.
This is nothing but the modified wave function in the pseudo position space \eqref{eq:xi_wf};
namely, the wave function in the pseudo-position space can be viewed as the conventional
Fourier transform of the auxiliary momentum space wave function.\footnote{%
It should be noted that this form is different from
the wave function in the position space of minimum uncertainty
\eqref{eq:position_wf_KMM} originally proposed by Kempf et al.\cite{Kempf:1994su}.}
We thus rename the variable $x$ as $\xi$ and look at the correction term.

\begin{figure}[tbh]
  \centering
  \includegraphics[scale=0.6]{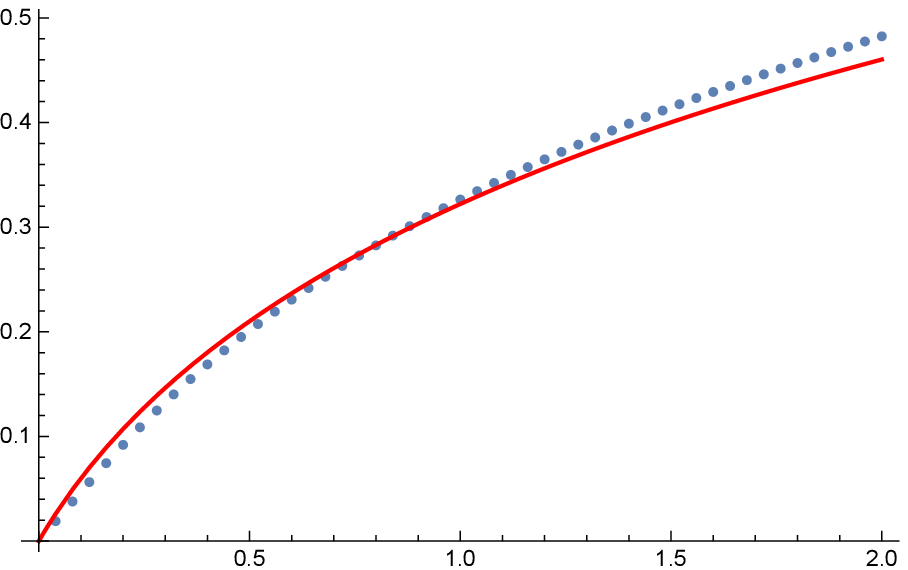}
\hspace{2em}
  \includegraphics[scale=0.6]{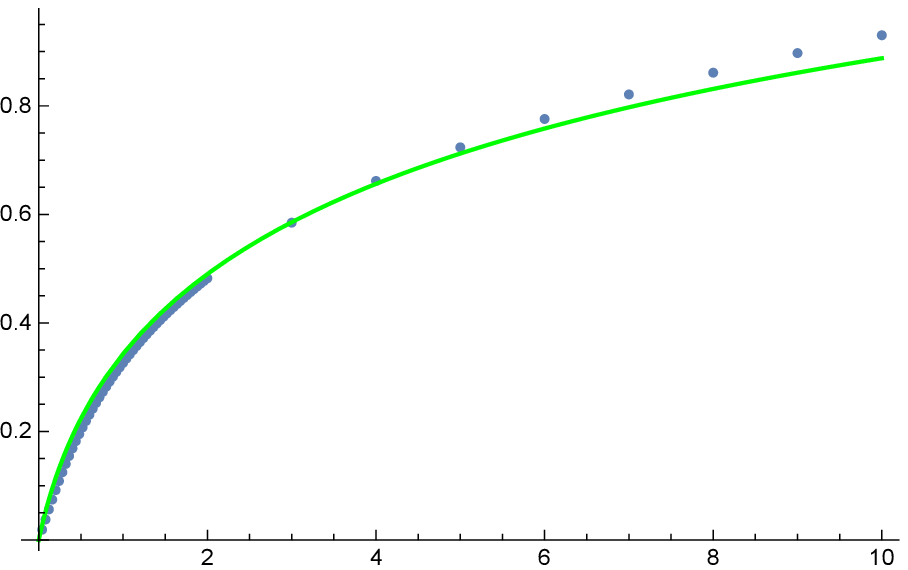}
  \caption{Fitting for the correction term $K(\beta)$
\eqref{eq:new_K0}
with $\alpha=1.39$ for $\beta\leq 2.0$ (Left) and
with $\alpha=1.48$ for $\beta\leq 10$ (Right).
Here $\sigma=\hbar=1$.}
\label{fig:fitting20}  
\end{figure}

Since $\psi(\xi)$ and $\psi(q)$ are usual Fourier pair, the Shannon entropies based
on the probability distributions from these wave functions satisfy
Bia\l ynicki-Birula and Mycielski type entropic uncertainty relation
\begin{align}
  \tilde{H}(\xi|\psi)+ \tilde{H}(q|\psi) \geq \log (\pi e) \,.
\end{align}
where $\tilde{H}(q|\psi)=-\vev{\log \tilde{\rho}(q)}_{\tilde{\rho}(q)}$ is calculated by use of a probability distribution $\tilde{\rho}(q)$.
We may relate this with the distribution in $p$ space by $\rho(p)dp=\tilde{\rho}(q)dq$,
which implies $\tilde{\rho}(q)=(1+\beta p^2)\rho(p)$.
Here $\rho(p)dp$ and $\tilde{\rho}(q)dq$ are normalized in the intervals $-\infty < p < \infty$
and $-q_0 \leq q \leq q_0$ respectively.
Thus, the Shannon entropy for $\rho(p)$ can be evaluated as
\begin{align}
  \tilde{H}(p|\psi)=&-\int_{-\infty}^{\infty}dp \, \rho(p) \log \rho(p)
=-\int_{-\infty}^{\infty}dp \, \rho(p) \big( \log \tilde{\rho}(q) - \log (1+\beta p^2) \big)
\nn\\=&
\tilde{H}(q|\psi)+\vev{\log (1+\beta p^2)}_{\rho(p)} \,.
\end{align}
By introducing bins for $\xi$ and $p$ spaces, we obtain the refined entropic
uncertainty relation with a correction term \cite{Rastegin:2016xeb},
\begin{align}
\label{eq:EUR2}
  H(\xi|\psi) + H(p|\psi) \geq \log \bigg( \frac{\pi e \hbar}{\delta x \delta p} \bigg)
+\vev{\log \big( 1+\beta p^2\big)}_\rho \,.
\end{align}
We may identify the extra term on the right hand side as the correction term $K$
\begin{align}
  K(\beta)=\alpha \vev{\log \big( 1+\beta p^2\big)}_\rho \,,
\label{eq:new_K0}
\end{align}
where $\alpha$ is an overall normalization factor.
Note that the correction term here
is \textit{state dependent} and
 is not exactly what we seek for.
It also does not depend on the bin size.
However,  we may expect that the correction term 
evaluated with the Gaussian wave function.
By tuning $\alpha$, 
we fit the numerical data of the Shannon entropy by $K(\beta)$ numerically
in Fig.~\ref{fig:fitting20}.
As seen, it provides a rather nice correction term.

\begin{figure}[tbh]
  \centering
  \includegraphics[scale=0.6]{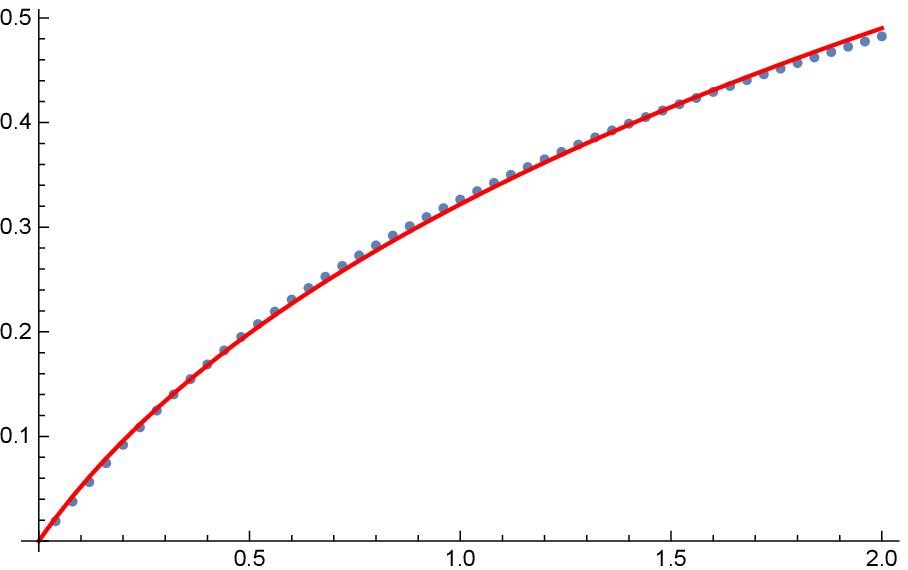}
\hspace{2em}
  \includegraphics[scale=0.6]{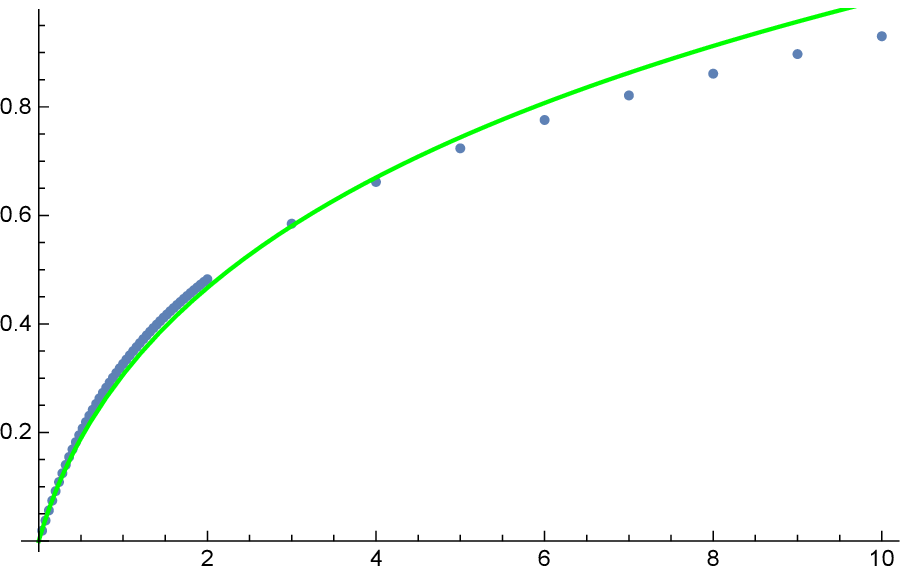}
  \caption{Fitting for the correction term $\tilde{K}(\beta)$
\eqref{eq:new_K}
with $\alpha=1.161$ for $\beta\leq 2.0$ (Left) and
with $\alpha=1.106$ for $\beta\leq 10$ (Right).
Here $\sigma=\hbar=1$.}
\label{fig:fitting2}  
\end{figure}

Though $K(\beta)$ can only be evaluated numerically, there is an analytic expression for
a correction term.
From Jensen's inequality, we have
$\log \big( 1+\beta \vev{p^2}_{\rho}\big) \geq \vev{\log (1+\beta p^2)}_{\rho}$.
The left hand side, $\tilde{K}(\beta)=\alpha \log \big( 1+\beta \vev{p^2}_{\rho}\big)$,
can be evaluated analytically as
\begin{align}
  \tilde{K}(\beta)=
\alpha
\log \bigg[
1+\frac{1}{\sqrt{\beta}}
\sqrt{ -1 + \frac{\hbar \sqrt{\pi}}{\sigma} F(\beta)^2 }
\bigg]\,,
\label{eq:new_K}
\end{align}
where a normalization factor $\alpha$ is introduced again.
Since $\tilde{K}(\beta) \geq K(\beta)$, this correction term would
overestimate the actual bound, but it turns out that
$\tilde{K}(\beta)$ fits the numerical result nicely,
especially for small $\beta$, as shown in Fig.~\ref{fig:fitting2}.

\section{Conclusion and discussion}
\label{sec:conclusion}

In this paper, we explore the correction to the entropic uncertainty relation
due to generalized uncertainty principle for small value of the deformation parameter
$\beta$.
The Gaussian wave function is used since it tends to saturate the
standard bound of entropic uncertainty relation for sufficiently fine bin
sizes.
The correction due to GUP is represented as a deformation of the wave
functions,
and the Shannon entropies for these $\beta$ deformed probability
distributions are calculated numerically.
We first fit the result by some simple functions of $\beta$;
for small $\beta$, we obtain 
$K(\beta;\delta \xi, \delta p) \sim
\beta^{0.7}$ as an approximated correction term.
a state dependent correction term 
based on an auxiliary momentum variable \cite{Pedram:2016gps, Rastegin:2016xeb},
for the Gaussian wave function.

We have used a pair of Gaussian wave functions in position and momentum representation.
In the undeformed case ($\beta=0$), the EUR saturates the bound \eqref{eq:EUR1}
for sufficiently fine bin sizes.
The deformation alters both momentum and position (or more precisely, pseudo-position)
probability distributions.
The behavior of the Shannon entropies for these distributions are consistent with consideration
based on GUP; namely,
the correction term $K(\beta;\delta \xi, \delta p)$ appears as a monotonically increasing function of $\beta$, 
which is anticipated from $\Delta x \Delta p$ in GUP.
We also consider an auxiliary momentum variable $q$ that satisfies the standard Heisenberg
algebra with $x$ \cite{Pedram:2016gps} and then the wave functions are related by the usual Fourier
transform. Putting $q$ back to the physical momentum, we find that the relation between 
the wave functions of physical momentum and pseudo-position space is naturally reproduced.
This procedure also provides a state-dependent correction term. 
This correction term is evaluated with Gaussian wave function,
we find that this correction term fits the numerical result very well.

So far, the corrections to the bound is investigated numerically.
It is obviously important to explore the analytic form of the
correction to understand the property of quantum gravity
through entropic entropy.
For example, in a finite dimensional system,
an explicit map between a variance type uncertainty relation
and an entropic type of it have been constructed \cite{Li-Qiao}.
It is then intriguing to investigate a similar type of mappings in our case.\footnote{%
We are informed that there has been a work that demonstrates an explicit connection
between the entropic and variance based uncertainty relations \cite{Yichen1}.}
Out method and result serve a first hint toward this direction.

GUP implies various modification of the consequence of standard theory of gravity:
for example the thermodynamic properties of black holes \cite{Adler:2001vs, Fabio1,Fabio2,Fabio3, Gangopadhyay:2015zma},
inflationary cosmology \cite{Ali:2015ola}, and
the entropic bound \cite{Wang:2012ku}.
It should be interesting to reformulate these kinds of modification in terms of
the deformed entropic uncertainty relation.

At the final stage of the first version of this work,
we noticed a paper on entropic uncertainty relation
with a minimal length \cite{Pedram:2016gps}
which addresses a similar question but uses a different approach and
another paper \cite{Rastegin:2016xeb} appeared later to
to analyze EUR based on \cite{Pedram:2016gps}.
In Sec.~\ref{sec:using-auxil-moment}, we discuss their formulation in our setup.
A related work \cite{Abdelkhalek:2016nyn} showed up too, which deals with the optimal
bound for both variance and entropic uncertainty relations in the presence of
minimal length.
Compared to ours, their paper argues more detailed bound for the sum of entropy functions,
but they use a special binning for $p$ space (we take $p$ as physical momentum and set up
a natural binning for $p$).

\section*{Acknowledgment}
\label{sec:acknowledgement}
This work is supported in part by
the Ministry of Science and Technology (MOST) of Taiwan
under the grant No.
~105-2112-M-033-003
~103-2811-M-033-004,
103-2119-M-007-003, and
102-2112-M-033-006-MY3,
and the National Center for Theoretical Sciences.


\end{document}